\documentclass[twocolumn,showpacs,preprintnumbers,amsmath,amssymb]{revtex4}

\usepackage{graphicx,epsfig}% Include figure files
\usepackage{dcolumn}% Align table columns on decimal point
\usepackage{bm}% bold math
\begin{document}

\title{Recurrence and P\'{o}lya number of general one-dimensional random walks}
\author{Xiao-Kun Zhang, Jing Wan, Jing-Ju Lu and Xin-Ping Xu}
\email{xuxinping@suda.edu.cn}
\affiliation{%
School of Physical Science and Technology, Soochow University,
Suzhou 215006, China
}%
\date{\today}
\begin{abstract}
The recurrence properties of random walks can be characterized by
P\'{o}lya number, {\it i.e.}, the probability that the walker has
returned to the origin at least once. In this paper, we consider
recurrence properties for a general 1D random walk on a line, in
which at each time step the walker can move to the left or right
with probabilities $l$ and $r$, or remain at the same position
with probability $o$ ($l+r+o=1$). We calculate P\'{o}lya number
$P$ of this model and find a simple expression for $P$ as,
$P=1-\Delta$, where $\Delta$ is the absolute difference of $l$ and
$r$ ($\Delta=|l-r|$). We prove this rigorous expression by the
method of creative telescoping, and our result suggests that the
walk is recurrent if and only if the left-moving probability $l$
equals to the right-moving probability $r$.
\end{abstract}
 \maketitle
Random walk is related to the diffusion models and is a
fundamental topic in discussions of Markov processes. Several
properties of (classical) random walks, including dispersal
distributions, first-passage times and encounter rates, have been
extensively studied. The theory of random walk has been applied to
computer science, physics, ecology, economics, and a number of
other fields as a fundamental model for random processes in
time~\cite{rn1,rn2,rn3,rn4}.

An interesting question for random walks is whether the walker
eventually returns to the starting point, which can be
characterized by P\'{o}lya number, {\it i.e.}, the probability
that the walker has returned to the origin at least once during
the time evolution. The concept of P\'{o}lya number was proposed
by George P\'{o}lya, who is a mathematician and first discussed
the recurrence property in classical random walks on infinite
lattices in 1921~\cite{rn5,rn6}. P\'{o}lya pointed out if the
number equals one, then the walk is called recurrent, otherwise
the walk is transient because the walker has a nonzero probability
to escape~\cite{rn7}. As a consequence, P\'{o}lya showed that for
one and two dimensional infinite lattices the walks are recurrent,
while for three dimension or higher dimensions the walks are
transient and a unique P\'{o}lya number is calculated for
them~\cite{rn8}. Recently, M. \v{S}tefa\v{n}\'{a}k {\it et al.}
extend the concept of P\'{o}lya number to characterize the
recurrence properties of quantum walks~\cite{rn9,rn10,rn11}. They
point out that the recurrence behavior of quantum walks is not
solely determined by the dimensionality of the structure, but also
depend on the topology of the walk, choice of coin operators, and
the initial coin state, etc~\cite{rn9,rn10,rn11}. This suggests
the P\'{o}lya number of random walks or quantum walks may depends
on a variety of ingredients including the structural
dimensionality and model parameters.

In this paper, we consider recurrence properties for a general
one-dimensional random walk. The walk starts at $x=0$ on a line
and at each time step the walker moves one unit towards the left
or right with probabilities $l$ and $r$, or remain at the same
position with probability $o$ ($l+r+o=1$). This general random
walk model has some useful application in physical or chemical
problems, and some of its dynamical properties requires a further
study. Previous studies of one-dimensional random walk focus on
the simple symmetric case where the walker moves to left and right
with equal probability ($l=r=1/2$)~\cite{rn10}. For instance,
P\'{o}lya showed that the symmetric random walk is recurrent and
its P\'{o}lya number equals to 1~\cite{rn12,rn13}. However,
recurrence properties of this general random walk defined here are
still unknown. As a consequence, we will calculate the P\'{o}lya
number for this general random model and discuss its recurrence
properties. We will try to derive an explicit expression for
P\'{o}lya number, and reveal its dependence on the model
parameters $l$, $r$ and $o$.

P\'{o}lya number of random walks can be expressed in terms of the
return probability $p_0(t)$~\cite{rn12,rn10}, {\it i.e.}, the
probability for the walker returns to its original position $x=0$
at step $t$,
\begin{equation}\label{eq1}
P=1-\frac{1}{\sum_{t=0}^{\infty}p_0(t)}.
\end{equation}
Hence, the recurrence behavior of random walk is determined solely
by the infinite summation of return probabilities. It is evident
that if the summation of return probabilities diverges the walk is
recurrent ($P=1$), and if the summation converges the walk is
transient ($P<1$). To calculate the P\'{o}lya number, it is
crucial to obtain the return probabilities. In the following, we
will calculate the return probabilities for our general random
walk model.

The return probability $p_0(t)$ can be obtained using the
trinomial coefficients of $(l+o+r)^t$. Considering an ensemble of
random walks after $t$ steps, in which the walker has $L$ steps
moving left, $R$ steps moving right and $O$ steps remaining at the
same position, then the probability for such random walks is
$\frac{t!}{O!L!R!}o^Ol^Lr^R$ ($l+o+r=1$, $L+O+R=t$). Since the
walker's position $x$ is only dependant on the difference of
right-moving steps $R$ and left-moving steps $L$, $x=R-L$,
returning to the original position $x=0$ requires $R=L$.
Therefore, the ensemble of random walks returning to $x=0$
involves sum over all possible $O$ subject to the constraints
$R=L$ and $R+L+O=t$. Because $R+L$ is an even number, $t$ and $O$
must have the same parity. Here, we suppose $t=2n$, $O=2i$ for
even $t$ and $O$, and $t=2n+1$, $O=2i+1$ for odd $t$ and $O$ ($i$
and $n$ are nonnegative integers, and $i\leq n$). We calculate the
return probability for even $t$ and odd $t$ separately. For even
$t$, the return probability is given by,
\begin{equation}\label{eq2}
p_0(t)|_{t=2n}=\sum_{i=0}^{n}\frac{(2n)!}{(2i)!(n-i)!(n-i)!}o^{2i}l^{n-i}r^{n-i},
\end{equation}
where $t=2n$, $O=2i$, $R=L=(t-O)/2=n-i$ are used in the above
equation. Analogously, for odd $t$, the return probability is
given by,
\begin{equation}\label{eq3}
p_0(t)|_{t=2n+1}=\sum_{i=0}^{n}\frac{(2n+1)!}{(2i+1)!(n-i)!(n-i)!}o^{2i+1}l^{n-i}r^{n-i}.
\end{equation}

The infinite summation of return probabilities $S$ can be
determined by the sum of $p_0(2n)$ and $p_0(2n+1)$,
\begin{equation}\label{eq4}
S=\sum_{t=0}^{\infty}p_0(t)=\sum_{n=0}^{\infty}\Big(p_0(t)|_{t=2n}+p_0(t)|_{t=2n+1}\Big).
\end{equation}

In order to get a simple expression for $S$, we define
$\Delta=|r-l|$, thus $lr=\big((1-o)^2-\Delta^2\big)/4$.
Substituting this relation into Eq.~(\ref{eq4}), we get
\begin{equation}\label{eq5}
\begin{array}{ll}
S&=\displaystyle{\sum_{n=0}^{\infty}}\Big(p_0(t)|_{t=2n}+p_0(t)|_{t=2n+1}\Big) \\
&=\displaystyle{\sum_{n=0}^{\infty}}\Big(\sum_{i=0}^{n}\frac{(2n)!}{(2i)!(n-i)!(n-i)!}o^{2i}\big(\frac{(1-o)^2-\Delta^2}{4}\big)^{n-i}\\
&\
 +\displaystyle{\sum_{i=0}^{n}}\frac{(2n+1)!}{(2i+1)!(n-i)!(n-i)!}o^{2i+1}\big(\frac{(1-o)^2-\Delta^2}{4}\big)^{n-i}\Big)\\
&=\displaystyle{\sum_{n=0}^{\infty}}\frac{(2n)!}{(n!)^2}\big(\frac{(1-o)^2-\Delta^2}{4}\big)^n
\\
& \times \Big(\ _2F_1(-n,-n,1/2,\frac{o^2}{(1-o)^2-\Delta^2})+ \\
& (2n+1)o\ _2F_1(-n,-n,3/2,\frac{o^2}{(1-o)^2-\Delta^2})\Big),
\end{array}
\end{equation}
where $_2F_1(a,b,c,z)$ is the Gauss Hypergeometric function. $S$
can be further simplified, for the sake of clarity, we first
consider the case $o=0$. When $o=0$ the Hypergeometric function
equals to 1, $S$ can be simplified as,
\begin{equation}\label{eq6}
S=\sum_{n=0}^{\infty}\frac{(2n)!}{(n!)^2}\Big(\frac{1-\Delta^2}{4}\Big)^n=\frac{1}{\Delta}.
\end{equation}
The last equality follows from the Taylor series expansion at
$z=0$ for the function $1/\sqrt{1-4z}$.

For $o>0$, we find that $S$ also equals to $1/\Delta$. This result
is surprising because $S$ does not depend on the remaining
unmoving probability $o$. This suggests that, for all $o$ and
$\Delta$, Eq.~(\ref{eq5}) can be simplified as,
\begin{equation}\label{eq7}
S=\frac{1}{\Delta}, \ \ \ \ \  \forall \ \  0<o, \Delta\leq1,
o+\Delta\leq1.
\end{equation}

It is difficult to simplify Eq.~(\ref{eq5}) or prove
Eq.~(\ref{eq7}) using the usual mathematical methods. Here, in the
appendix, we prove this rigorous expression (\ref{eq7}) by the
method of creative telescoping. The method of creative
telescoping~\cite{rn14,rn15,rn16} is an algorithm to compute
hypergeometric summation, definite integration, and prove
combinatorial identity. Using this method, we transfer $S$ to the
solution of a partial differential equation (See the proof in the
appendix).

The P\'{o}lya number in Eq.~(\ref{eq1}) can be written as,
\begin{equation}\label{eq8}
P=1-\frac{1}{S}=1-\Delta.
\end{equation}
Consequently, we find a simple explicit expression for P\'{o}lya
number, which is solely determined by the absolute difference of
$l$ and $r$, $\Delta=|l-r|$.

\begin{figure}
\scalebox{0.25}[0.25]{\includegraphics{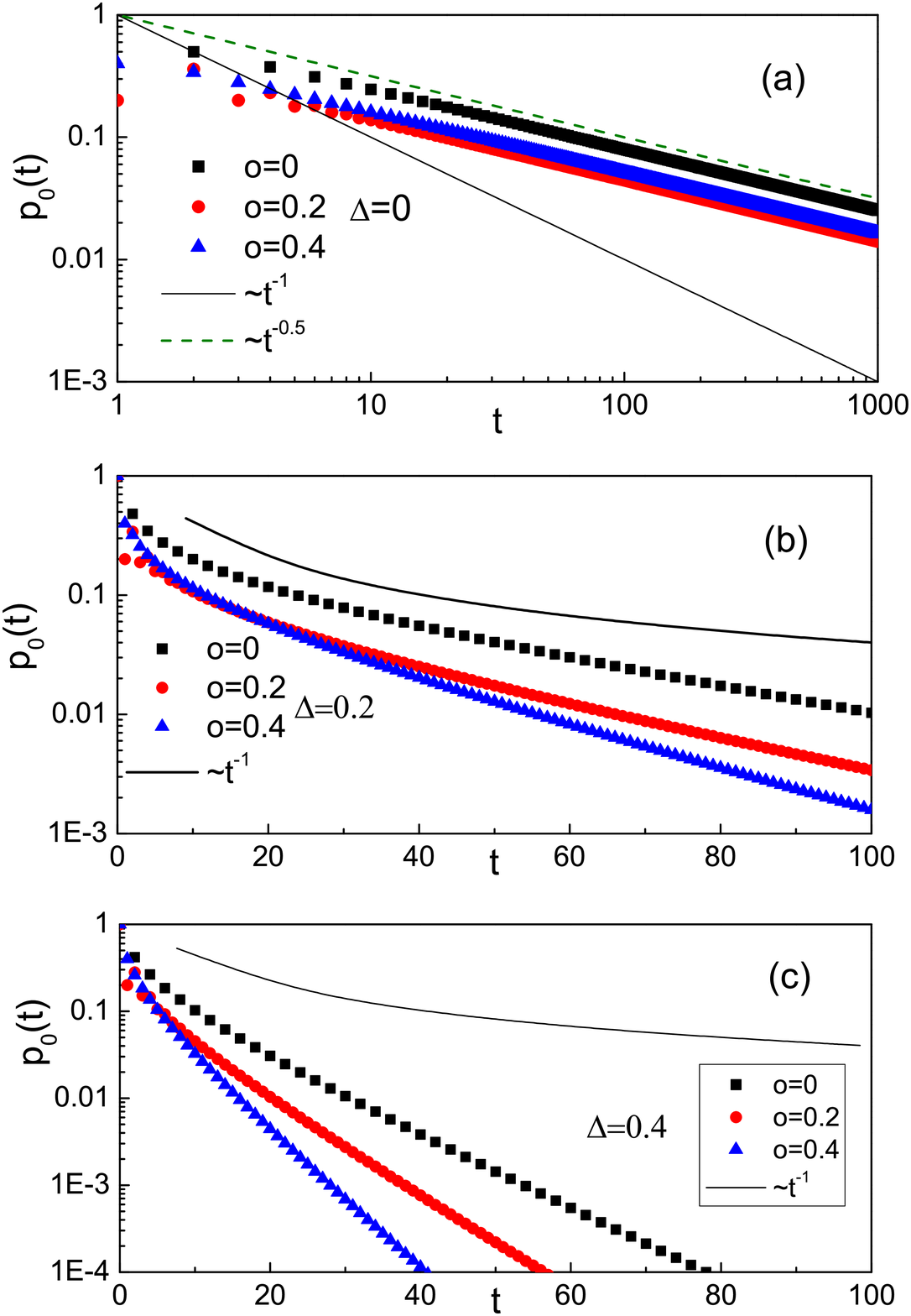}} \caption{(Color
online) Return probability $p_0(t)$ as a function of step $t$ for
$\Delta=0$ (a), $\Delta=0.2$ (b) and $\Delta=0.4$ (c). For each
value of $\Delta$, we plot $p_0(t)$ vs $t$ for $o=0$ (black
squares), $o=0.2$ (red dots) and $o=0.4$ (blue triangles). The
critical decay for convergence $p_0(t)\sim t^{-1}$ are also
plotted in the figure. $p_0(t)$ shows a power-law decay $t^{-0.5}$
for $\Delta=0$ (See (a)), and $p_0(t)$ exhibits exponential decay
for $\Delta>0$ (See (b) and (c)). It should be pointed out that
for the case $o>0$, $p_0(t)$ is nonzero at all values of $t$,
while $p_0(t)$ is zero at odd $t$ for $o=0$.
 \label{fg1}}
\end{figure}

According to Eq.~(\ref{eq8}), P\'{o}lya number $P$ equals to 1 for
$\Delta=0$. This suggests that the walk is recurrent if and only
if the left-moving probability $l$ equals to the right-moving
probability $r$. Our result is consistent with previous conclusion
that one-dimensional symmetric random walk ($l=r=1/2$) is
recurrent. Our result also indicates that the infinite summation
of return probabilities $S$ diverges for $\Delta=0$ and converges
for $\Delta\neq0$. To verify this point, we plot the return
probability $p_0(t)$ as a function of step $t$ in Fig.~\ref{fg1}.
We find that $p_0(t)$ is a power-law decay as $p_0(t)\sim t^{-0.5}$
for $\Delta=0$ (See Fig.~\ref{fg1} (a) in the log-log plot) and
exponential decay for $\Delta\neq0$ (See Fig.~\ref{fg1} (b), (c)
in the log-linear plot). Since $p_0(t)$ for $\Delta=0$ decays
slower than $t^{-1}$ and decays faster than $t^{-1}$ for
$\Delta\neq0$, the infinite summation $S$ diverges for $\Delta=0$
and converges otherwise. Particularly, by means of Stirling's
approximation $n!\approx\sqrt{2\pi n}(n/e)^n$ for $o=0$, we find
an asymptotic form for the return probability in Eq.~(\ref{eq6}):
$p_0(t)\approx\sqrt{\frac{2}{\pi t}}(1-\Delta^2)^{t/2}$ for even
$t$ and $p_0(t)=0$ for odd $t$. For a certain value of $\Delta>0$,
the decay behavior of $p_0(t)$ seems different for different
values of $o$ (See Fig.~\ref{fg1} (b), (c)). However, the
summations of $p_0(t)$ for different $o$ are identical and equal
to $1/\Delta$. This result is some what unexpected and we provide
a strict proof in the appendix.

In summary, we have studied recurrence properties for a general 1D
random walk on a line, in which at each time step the walker can
move to the left or right with probabilities $l$ and $r$, or
remain at the same position with probability $o$ ($l+r+o=1$). We
calculate P\'{o}lya number $P$ of this model for the first time,
and find a simple explicit expression for $P$ as, $P=1-\Delta$,
where $\Delta$ is the absolute difference of $l$ and $r$
($\Delta=|l-r|$). We prove this rigorous relation by the method of
creative telescoping, and our result suggests that the walk is
recurrent if and only if the left-moving probability $l$ equals to
the right-moving probability $r$.

We thank Armin Straub and Dr. Koutschan for useful discussions.
This work is supported by National Natural Science Foundation of
China under project 10975057, the new Teacher Foundation of
Soochow University under contracts Q3108908, Q4108910, and the
extracurricular research foundation of undergraduates under project KY2010056A.

\appendix
\section{The method of creative telescoping (MCT)}
The method of creative telescoping, also known as Zeilberger's
algorithm~\cite{rn14,rn15,rn16}, is a powerful tool for solving
problem involving definite integration and summation of
hypergeometric function. Suppose we are given a certain holonomic
function of two variables $F(z, n)$ ($n\in Integers$, $z\in
Reals$), and it is required to prove that the summation of $F(z,
n)$ over $n$ equals to $f(z)$,
\begin{equation}\label{eq9}
\sum_n F(z,n)=f(z).
\end{equation}
The basic idea of creative telescoping algorithm is to find a
linear recurrence equation for the summands $F(z,n)$. This could be done by
constructing a differential operator $\hat{L}$ with coefficients
being polynomials in $z$, and a new function $G(z,n)$ satisfying,
\begin{equation}\label{eq10}
\hat{L}(z) F(z,n)=G(z,n+1)-G(z,n).
\end{equation}
Thus $\hat{L}(z)$ operating on the summation $\sum_nF(z, n)$ is
determined by the difference of upper bound and lower bound
$G_0(z)=G(z,n_{max})-G(z,n_{min})$. Then we just need to check
both sides of Eq.~(\ref{eq9}) satisfy recurrence equations:
$\hat{L}(z)\sum_nF(z,n)=G_0(z)$, $\hat{L}(z)f(z)=G_0(z)$,
and check Eq.~(\ref{eq9}) holds for some initial conditions.

Several algorithms for computing creative telescoping relations
have been developed in the past~\cite{rn17}. The main programs are
Zeilberger's Maple program and Mathematica program written by
Peter Paule and Markus Schorn~\cite{rn17,rn18,rn19}. Here, we use
the mathematical program to compute the creative telescoping
relation for our problem.
\section{Proof of $S=\frac{1}{\Delta}$ using MCT}
In this section, we prove $S=\frac{1}{\Delta}$ using the method of
creative telescoping (MCT). We use the Mathematica package
Holonomic Functions~\cite{rn17,rn20,rn21} to create a recurrence
relation for the summands $s_n(o,\Delta)$ in Eq.~(\ref{eq5}),
\begin{equation}\label{eq11}
\Big(2oD_o + \Delta D_{\Delta} + 1 + (S_n - 1)\frac{1}{\Delta}
D_{\Delta}\Big)s_n(o,\Delta)=0,
\end{equation}
where $D_o$, $D_{\Delta}$ are the partial differential operator
($D_o\equiv\partial /\partial o$, $D_{\Delta}\equiv\partial
/\partial \Delta$), $S_n$ is the shift operator satisfying
$S_nf(n)=f(n+1)$.

Summing over $n$ leads to,
\begin{equation}\label{eq12}
\Big(2oD_o + \Delta D_{\Delta} + 1\Big)S +
\displaystyle{\sum_{n=0}^{\infty}}(S_n - 1)\frac{1}{\Delta}
D_{\Delta}s_n(o,\Delta)=0.
\end{equation}
The second term in the above equation is a telescoping series, the
central terms are cancelled and only leave the last term and first
term. Noting that $\frac{1}{\Delta} D_{\Delta}s_n(o,\Delta)$ are
zero for $n=0$ and $n\rightarrow\infty$, the second term in
Eq.~(\ref{eq12}) equals to $0$. Hence the infinite summation of
return probabilities $S$ satisfies,
\begin{equation}\label{eq13}
\Big(2oD_o + \Delta D_{\Delta} + 1\Big)S=0.
\end{equation}
It is easy to check $\frac{1}{\Delta}$ also satisfies the above
partial differential equation. Combining with the initial
condition $S=\frac{1}{\Delta}$ for $o=0$ (See Eq.~(\ref{eq6})),
$S=\frac{1}{\Delta}$ holds for all $o$ and $\Delta$.

\end{document}